\documentstyle[12pt]{article}
\newcommand{\rf}[1]{(\ref{#1})}
\newcommand{\ra}{\rightarrow}
\newcommand{\beq}{\begin{equation}}
\newcommand{\eeq}{\end{equation}}
\newcommand{\bea}{\begin{eqnarray}}
\newcommand{\eea}{\end{eqnarray}}

\renewcommand{\l}{\lambda}

\newcommand{\Ar}{{\cal A}}
\newcommand{\Per}{{\cal P}}
\newcommand{\s}{\sigma}
\newcommand{\n}{\nu}
\newcommand{\m}{\mu}

\newcommand{\oh}{\frac{1}{2}}

\newcommand{\dg}{\dagger}
\newcommand{\non}{\nonumber}
\begin{document}

\addtolength{\baselineskip}{0.20\baselineskip}
\hfill  IFUP-TH 54/92

\hfill hep-lat/9301021

\hfill   November 1992
\begin{center}

\vspace{36pt}
{\large \bf Charge Screening, Large-N, and the Abelian Projection
Model of Confinement}

\end{center}

\vspace{36pt}

\begin{center}
{\sl L. Del\/ Debbio}
\vspace{12pt}

Dipartimento di Fisica dell' Universit\`a, and I.N.F.N. \\
I-56100 Pisa, Italy

\vspace{12pt}

{\sl M. Faber}
\footnote{Supported in part by the Fonds zur F\"orderung der
wissenschaftlichen Forschung under Contract No. P7237-TEC.}

\vspace{12pt}

Institut f\"ur  Kernphysik \\
Wiedner Hauptstrasse 8-10 \\
A-1040 Wien, Austria \\

\vspace{12pt}
{\sl J. Greensite}
\footnote{Supported by the U.S. Department of Energy under
Grant No. DE-FG03-92ER40711.}
\vspace{12pt}

Physics and Astronomy Dept.\\
San Francisco State University\\
1600 Holloway Ave.\\
San Francisco, CA 94132 USA\\

\end{center}
\vfill
\newpage

\begin{center}
{\bf Abstract}
\end{center}
\vspace{12pt}

    We point out that the abelian projection theory of quark
confinement is in conflict with certain large-N predictions.
According to both large-N and lattice strong-coupling arguments,
the perimeter law behavior of adjoint Wilson loops at large scales
is due to charge-screening, and is suppressed relative to the area
term by a factor of $1/N^2$.  In the abelian projection theory, however,
the perimeter law is due to the fact that $N-1$ out of $N^2-1$
adjoint quark degrees of freedom are (abelian) neutral and unconfined;
the suppression factor relative to the area law is thus only $1/N$.
We study numerically the behavior of Wilson loops and Polyakov lines with
insertions of (abelian) charge projection operators, in maximal abelian
gauge.  It appears from our data
that the forces between abelian charged, and abelian neutral adjoint
quarks are not significantly different.  We also show via the lattice
strong-coupling expansion that, at least at strong couplings, QCD flux
tubes attract one another, whereas vortices in type II
superconductors repel.

\vfill
\newpage

\section{Introduction}

   Large-N arguments are a useful way to critique some of the quark
confinement mechanisms that have been proposed over the years.  For
example, the $Z_N$ fluxon mechanism \cite{fluxon} does not give
a string tension for adjoint quarks at any length scale.  This was
shown to be in contradiction with large-N factorization, which predicts
that: (i) the adjoint string tension $\sigma_A$
(from the confinement scale to the charge screening scale $L_s$) is
roughly twice the fundamental string tension $\sigma_F$;
and (ii) the distance $L_s$ where the adjoint flux tube
breaks, due to charge screening, goes to $\infty$ as $N \ra \infty$
\cite{GH}.  Existing Monte Carlo calculations appear to be generally
consistent with these large-N predictions \cite{MC}.
A quite different mechanism is the "dual superconductor"
idea, particularly as formulated by 't Hooft in abelian-projection gauges
\cite{'t Hooft}; this formulation has been widely discussed in recent
years [5-11]. It is obviously of interest to see whether
the abelian projection idea is in agreement with large-N predictions.

   At first sight, the abelian projection theory meets the large-N
criteria quite well:  adjoint quarks will indeed have a string tension
$\sigma_A \approx 2 \sigma_F$ from the confinement scale to some
intermediate distance, beyond which the adjoint quarks are unconfined
\cite{Rosenzweig}.  However, as discussed in section 3, there is still
a profound contradiction between large-N and abelian-projection
predictions,
namely, for an adjoint Wilson loop of area $\Ar$ and perimeter $\Per$ we
expect
at large-N the leading behavior

\beq
      W(C) = N^2 \left[e^{-\sigma \Ar -\m \Per} +
           {1 \over N^2} e^{-\m' \Per} \right]
\label{WN}
\eeq
while for abelian projection we find

\beq
      W(C) = N^2 \left[e^{-\sigma \Ar -\m \Per} +
           {1 \over N} e^{-\m' \Per} \right]
\eeq

  This may not seem like a very serious difference - just a $1/N$ vs.
$1/N^2$
suppression of the perimeter law term - but in fact it reveals something
very fundamental about the abelian projection mechanism:  In abelian
projection gauge, certain adjoint quark colors are unconfined not because
of
charge screening, but rather because they are neutral with respect to the
residual $U(1)^{N-1}$ gauge invariance, and this in turn leads to results
that differ with large-N.  The disagreement suggests certain
numerical tests of the abelian projection idea, based on the behavior of
Wilson loops and Polyakov lines with insertions of various abelian charge
projection operators.  It is found, in section 4, that these projection
operators do not make much difference so far as string tension and
screening distance are concerned, and that the force between quarks which
are neutral with respect to the abelian projection subgroup is about the
same as the force between quarks which are charged in the abelian subgroup.
These numerical results do not favor the abelian projection theory,
and add to the negative body of evidence already presented in ref.
\cite{Luigi,Win,Olejnik}.
We will comment on the apparently positive results presented in ref.
\cite{Hioki,Yots}.

   We will also comment, in the context of the lattice strong-coupling
expansion, on the force between QCD flux tubes.  Flux tubes in QCD have
often been likened to Abrikosov vortices in type II superconductors.
The force between vortices in type II superconductors is repulsive, due
to the negative surface energy of the vortices, so it is natural to ask
whether the force between QCD flux tubes is also repulsive.  In the next
section it will be shown that, at least at strong-couplings, the force
between QCD flux tubes is actually attractive.

\section{Charge Screening and Flux Coalescence}

    Strong-coupling calculations using the Kogut-Susskind Hamiltonian,
or based on the Heat-Kernel action, as well as Monte-Carlo
calculations in the scaling region with the Wilson action \cite{MC},
all give the result that the force between a quark and an antiquark, with
color charges in representation $R$ of the gauge group $SU(N)$, is
initially
proportional to the quadratic Casimir $C_R$ of the group representation.
For any representation of SU(N) there is a generalization of the concept
of triality in SU(3), known as the "N-ality" or, in the mathematical
literature, the "class" of the representation.  Another result of
strong-coupling \cite{GH} and Monte-Carlo \cite{MC} calculations
is that at some distance beyond the confinement scale, at a screening
distance denoted $L_s$, the force between quarks drops abruptly.
Beyond the screening distance, the force
becomes proportional to the smallest Casimir $C_L$ among all
representations
with the same N-ality as representation $R$.  This is known as
"charge-screening", and it has a simple physical explanation: As the
quark-antiquark pair separates and and the energy stored in the flux-tube
increases, it becomes energetically favorable to pair-create gluons.
These gluons bind to the quark and the antiquark and, although gluons
cannot change the N-ality of the quark charge, they can lower the
effective charge of the quark-gluon state to a representation $L$ of
the same N-ality, with the lowest possible Casimir $C_L$.
A particular example is the case of having a pair of quarks in the
adjoint representation.  The adjoint representation has N-ality = 0,
the same as a singlet.  A pair of gluons, binding to each quark, can
reduce the charge to a singlet, and therefore just break the flux-tube
between the adjoint quarks.  The mechanism is exactly the same as in
QCD with dynamical quarks, where a flux-tube is broken by quark-antiquark
production.  For this reason, in pure QCD, we expect an adjoint Wilson
loop to have an area law from the confinement scale up to a distance
$L_s$, after which it is screened to perimeter law.  This is indeed
what is seen in numerical experiments \cite{MC}.

   An important point, especially for the purposes of this article,
is that diagramatically the charge-screening process is always suppressed
by a factor of $1/N^2$ in pure QCD; this is true
whether the process is represented
by lattice strong-coupling diagrams, as shown in Fig. 1, or by
high-order Feynman diagrams.  Feynman diagrams can be classified
according to their associated powers of $1/N^2$, and diagrams
of leading order - the planar diagrams - satisfy the factorization property

\beq
      <(TrA)(TrB)> = <TrA><TrB>
\eeq
For this reason $\sigma_A=2\sigma_F$ in the large-N limit; it is also
the reason why charge screening must be a non-planar process.  Non-planar
diagrams, whether Feynman or lattice strong-coupling, are suppressed
(in pure QCD) by a factor of at least $1/N^2$ relative to the planar
diagrams.  Therefore, we conclude on these very general grounds that
for an adjoint Wilson loop

\beq
      W_A(C) = N^2 \left[ \exp[-2\sigma_F \Ar - \m \Per] + {1 \over N^2}
                  \exp[-\m' \Per]\right]
\label{WN1}
\eeq
In strong-coupling one can easily calculate where the perimeter behavior
of the second term takes over from the area behavior; for square
$L \times L$ loops this occurs for

\beq
         L > \left[ {\ln N \over \s_F} \right]^{1/2}
\label{Ls}
\eeq
which diverges as $N \ra \infty$.

   An effect which is very closely related to charge screening is the
phenomenon of flux coalescence.  This effect is relevant to the
question of whether the force between QCD flux tubes is repulsive,
as it is for Abrikosov vortices in type II superconductors, or
attractive.

    Let us consider two quark-antiquark pairs with the axes through
each pair parallel; these are represented by the two parallel $R \times T$
Wilson loops shown in Fig. 2, which are separated by a distance $L>>R$,
and which have the same orientation.
Denote the expectation value of the product of these loops by
$D(R,T,L)$.  Then the interaction energy between the loops, which
represents the interaction of the flux tubes between the quark-antiquark
pairs, is just

\beq
      V(L) = - \lim_{T \ra \infty} {\partial \over \partial T}
                \ln{\left[{D(R,T,L) \over D(R,T,\infty)}\right]}
\eeq
and

\beq
       D(R,T,L) = D(R,T,\infty) + RT(L+1)(A_a - A_b)
\label{A}
\eeq
where, to leading order
\beq
       D(R,T,\infty) = N^2 e^{-2 \s_F RT}
\eeq
The leading strong-coupling contributions to $A_a$ and $A_b$ are indicated
schematically in Fig. 3.

   We will calculate $V(L)$ using the heat-kernel lattice action

\bea
          e^S &=& \prod_p e^{S_p}
\non \\
          e^{S_p} &=& \sum_r d_r \chi_r(U_p) \exp[{-C_r \over N \beta}]
\eea
where the product extends over all oriented plaquettes $p$ and the
summation
runs over all inequivalent irreducible representations $r$ of dimension
$d_r$.  $\chi_r(U_p)$ is the character of the group element $U_p \in SU(N)$
in the representation $r$, and $C_r$ is the eigenvalue of the quadratic
Casimir operator in this representation.  For quarks in the fundamental
(defining) representation, the plaquette with character
$\chi^{\dagger}_r(U_p)$
in the middle of the tube in Fig. 2a can only be in the
symmetric ($r=s$) or antisymmetric ($r=a$) representations formed from
the product of two fundamental defining representations.
The interaction potential is then found to be

\beq
       V(L) = -(L+1) e^{-4 \s_F L} R \left[ {N-1 \over 2N} \exp[{1+N \over
N^2 \beta}] + {N+1 \over 2N} \exp[{1-N \over N^2 \beta}] - 1 \right]
\eeq
Expanding the exponents in a Taylor series, $V(L)$ to leading order
in $1/N$ is given by

\beq
       V(L) = - {1 \over 2 \beta^2}{1 \over N^2} R (L+1) e^{-4 \s_F L}
\eeq
which is an attractive potential at large $L>>R$.\footnote{In fact,
this potential is also attractive
at small $L$, given  $\s_F = C_F/(N \beta) > 1$, which is the
case at strong-couplings. But at $L<<R$, flux tube coalescence,
discussed below, is the dominant effect.}
Flux tubes therefore tend to attract one
another, by a process which can be viewed as glueball exchange.

   For separations $L<<R$ flux tubes do not only attract, they will
actually tend to coalesce into a single flux tube.  In this case

\beq
        D(R,T,L) = D(R,T,\infty) + D_c(R,T,L)
\eeq
where the leading contribution to $D_c$ is shown in Fig. 4.  In this
figure there is a single sheet of plaquettes, of area $R \times T$,
in a fundamental representation of N-ality N-2.  For $R << N L$ the
disconnected diagrams dominate, and the energy of the system is simply
the sum of the two disconnected flux tubes

\beq
         E_0 = 2 \s_F R = {N^2 -1 \over N^2 \beta} R
\eeq
whereas, if $R >> N L$, it is energetically favorable for the two
flux tubes to coalesce, as represented in Fig. 4, and the energy
is
\beq
         E_c = \s_a R + 2 \s_F L = {(N-2)(N+1) \over N^2 \beta}R
                        + {N^2 -1 \over N^2 \beta} L
\eeq
It is clear that $E_c<E_0$, i.e. coalescence is favored, for
$R>NL$.

   The calculation above can be easily generalized to the case
of $n$ quark-antiquark pairs, with axes nearby and aligned parallel
to one another.  Once again, for quark separations $L$ much less
than quark-antiquark separations $R$, it is energetically favorable
for the $n$ flux tubes between each quark-antiquark pair to coalesce
into a single tube, as shown in Fig. 5.  Its string tension is
determined by the eigenvalue of the quadratic Casimir operator of
the lowest dimensional representation $r$ of N-ality $(N-n)\mbox{mod}N$,
which is again a fundamental representation of dimension
$d_r = {N \choose n}$.
It is clear that coalescence
may result in considerable lowering of the energy of the system, compared
to the configuration of $n$ separate flux tubes between each
$q \overline{q}$ pair; the ratio between the energies of these two
different configurations being, for $R>>L$,

\beq
         r = {\sigma_L \over n \sigma_F} = {C_L \over n C_F}
           = 1 - {n-1 \over N-1}
\eeq
For n=2 in SU(3), this ratio is just $r=\oh$.

   The phenomenon of flux-tube coalesence implies that the force
between nearby QCD flux tubes is attractive, at least at strong
couplings, in contrast to the situation in type-II superconductivity.
{\it If} strong-coupling calculations are any guide, and {\it if}
confinement in QCD is analogous to (dual) superconductivity, then the
analogy
would presumably be to type I superconductivity.  More importantly,
both color charge screening and the attractive force between
flux tubes are due to non-planar, $1/N^2$ suppressed processes.  We will
now
consider whether the abelian projection theory is consistent with this
suppression factor.

\section{The Abelian Projection Theory}

    The abelian projection theory of quark confinement was put
forward in ref. \cite{'t Hooft}.
In this theory a gauge-fixing condition is first
chosen to break the SU(N) symmetry down to the Cartan subgroup
$U(1)^{N-1}$. Monopoles are then identified with singularities in the
gauge-fixing condition, and condensation of these monopoles is
invoked to explain confinement of particles charged with respect to
the $U(1)^{N-1}$ subgroup.  The confinement mechanism is analogous
that of compact QED, with gluons associated with the Cartan subalgebra
playing the role of the photon field which forms the flux tube.
These "diagonal" gluons are uncharged with respect to the residual
$U(1)^{N-1}$ symmetry; all other gluons are charged with respect
to the residual symmetry, and, like quarks, are confined by flux tubes.

   Consider the force between two quarks with charges in adjoint
representation, in the abelian projection theory.  For simplicity
(and because it makes little difference in the large-N limit), take
the group to be $U(N)$ rather than $SU(N)$, so the abelian subgroup
is $U(1)^N$.  Let $g_i$ denote an abelian charge of magnitude $g$
in the i-th $U(1)$ subgroup.  A quark in the fundamental (defining)
representation has $N$ color degrees of freedom, and from
the abelian-projection point of view,  each of these degrees of
freedom corresponds to a quark $Q^i_F$ with a different abelian charge
$g_i$, $1\le i \le N$.  Likewise, the $N^2$ degrees of freedom of
a quark in the adjoint representation can be be grouped according
to abelian charge:  there are $N(N-1)$ quarks $Q_A^{ij}$, $i \ne j$
with charge ($g_i,-g_j$), and there are $N$ quarks $Q_A^{ii}$ which
are {\it neutral} with respect to the $U(1)^N$ subgroup.

   Since the flux-tube between static adjoint quarks is neutral
with respect to the abelian subgroup in the abelian projection picture
(flux tubes are formed from the "photon" fields),
the charges of the quarks themselves must be correllated in order to
have a neutral composite state.  The quark-content of the composite state,
in an abelian-projection gauge, must then be

\beq
         Tr Q[x_1] Q[x_2]  = \sum_{i \ne j}Q^{ij}[x_1]Q^{ji}[x_2]
+ \sum_i Q^{ii}[x_1]Q^{ii}[x_2]
\label{adcharge}
\eeq
Consider the $N(N-1)$ contributions to the first sum.  Each contribution
represents two quarks with equal and opposite charges ($g_i,-g_j$) and
($-g_i,g_j$) respectively.  Since each quark is charged in two different
abelian subgroups, there will be two flux tubes between them, each with
string tension $\s_F$.  Thus the net string tension for these quarks
is $\s_A = 2 \s_F$, in agreement with large-N.  The $N$ quarks of the
second sum, however, are neutral with respect to the confining abelian
$U(1)^N$ subgroup; these neutral quarks are unconfined.  For an adjoint
Wilson loop of area $\Ar$ and perimeter $\Per$, we therefore expect

\beq
      W(C) = N(N-1) \exp[-2 \s_F \Ar - \m P] + N \exp[- \m \Per]
\label{UN}
\eeq
for the following reason:  Given that the abelian "charged" gluons
are confined and the flux tube is neutral, the quarks cannot exchange
charge between them, and a quark can only change its abelian charge
by emission and reabsorbtion of a charged gluon.  Even allowing for such
virtual processes, the abelian
charge of each quark + gluon-cloud cannot change.
Thus the color sum in a Wilson loop can be expressed as a sum of loops,
each representing a particular ($g_i,-g_j$) charge running around the loop.
The $N(N-1)$ charged adjoint quarks ($i\ne  j$) contribute to the first
term in \rf{UN}, and the $N$ neutral, unconfined quarks contribute to the
second term.

   The expression \rf{UN} above for the adjoint Wilson loop is in
disagreement with the large-N prediction.
In the large-N analysis, all quark colors are on the same footing,
and the perimeter term comes about through charge-screening, and not
because any subset of quark charges is oblivious to the confining force.
More importantly, according to large-N, the coefficient of the perimeter
term is $O(1)$, rather than $O(N)$ as in the abelian projection theory.

   To make the same point in a slightly different way, consider
the trace of an adjoint Wilson loop in $SU(N)$

\bea
        Tr_A \left[ P\exp[i\oint A^a L_a] \right]
   &=& \sum_m Tr_A \left[ |m)(m| P\exp[i\oint A^a L_a] \right]
\non \\ &=& \sum_m (m| P\exp[i\oint A^a L_a] |m)
\eea
where $L_a$ are the group generators and sum is over all members $|m)$
of the multiplet.  The confining force can only come from coupling
of the adjoint quarks to the gluons in the Cartan subalgebra, with
generators denoted $H_i$ ($i=1,...,N$).  Therefore, in an abelian
projection
gauge such as the maximal abelian gauge, following the reasoning of
ref. \cite{Smit},

\bea
       <Tr_A\left[ P\exp[i\oint A^a L_a] \right]> &\approx&
            <Tr_A \left[ \exp[i\oint A^i H_i] \right]>
\non \\
  &=& \sum_{m=1}^{N^2-1}<(m|\exp[i\oint A^i H_i] |m)>
\non \\
  &=& \sum_{m=1}^{N^2-1} <\exp[i\oint A^i \l^{(m)}_i]>
\label{JS}
\eea
where the $\l_i^{(m)}$ are eigenvalues
\beq
         H_i |m) = \l^{(m)}_i |m)
\eeq
Then, making use of the fact that the multiplicity of zero-weight
(all $\l_i=0$) states in the adjoint representation is $N-1$, we
have

\beq
      W(C) \approx \sum_{ \begin{array}{c}
       m=1 \\ \mbox{(non-zero} \\ \mbox{weight states)} \end{array}
}^{N(N-1)}
                    e^{-\s^{(m)} \Ar - \m^{\prime} \Per} ~~~~~+~~~~~ (N-1)
\eeq
Ignoring the coupling of adjoint quarks to gluons which are not in the
Cartan subalgebra,
which was the approximation made in eq. \rf{JS}, has the effect of dropping
self-energy (perimeter law) contributions to the zero-weight and
underestimating them for the non-zero
weight states.  If these contributions are included perturbatively,
the final answer would be

\beq
      W(C) \approx \sum_{ \begin{array}{c}
       m=1 \\ \mbox{(non-zero} \\ \mbox{weight states)} \end{array}
}^{N(N-1)}
                    e^{-\s^{(m)} \Ar - \m \Per}  ~~~~~+~~~~~
        (N-1) e^{-\m \Per}
\eeq
Again, the deconfined term is $O(N)$, rather than $O(1)$ as expected
by large-N arguments.

\section{Numerical Tests}

   The fact that the abelian projection theory disagrees with large-N
predictions should not be too surprising.  As noted above, the
deconfinement of adjoint loops according to the abelian projection
is simply due to the fact that $N-1$ out of
$N^2-1$ adjoint quarks are neutral with respect to the $U(1)^{N-1}$
subgroup, and are therefore unconfined in the abelian projection theory.
This gives a suppression factor of only $1/N$ to the unconfined
contribution.
Charge-screening, which involves pair-creating gluons that bind to
the adjoint quarks, is a completely different mechanism, and just gives
the usual non-planar suppression factor of $1/N^2$.  This disagreement
doesn't necessarily mean that the abelian projection theory is wrong;
it could be the large-N arguments that have somehow gone astray.  But
such a clear difference in the two approaches does suggest a simple
numerical test.

   Let us consider the case of $N=2$, with $H=L_3$ to be the
generator of the Cartan-subalgebra.  Then eq. \rf{JS} becomes
\bea
   W(C) &\approx& \sum_{m=-1}^1 <\exp[im\oint A^3]>
\non \\
        &=& 2 e^{\s^{(1)} \Ar} + 1
\eea
or, correcting for the self-energy (perimeter) contributions

\beq
   W(C) = 2 e^{\s^{(1)} \Ar - \m \Per} + e^{-\m \Per}
\eeq
In other words, the abelian-projection prediction is that,
in an abelian projection gauge

\bea
     <(1|P\exp[i\oint A]|1)> &=& <(-1|P\exp[i\oint A]|-1)>
  = e^{-\s^{(1)} \Ar - \m \Per}
\non \\
     <(0|P\exp[i\oint A]|0)> &=& e^{- \m \Per}
\eea
The inner product $<(0|...|0)>$ can be thought
of as creation of a quark-antiquark pair with zero abelian charge,
which run around the loop. The only way to change the abelian charge of
the pair would be for the quarks to exchange a
gluon with non-zero abelian charge.  However,
this process should be suppressed at the confinement scale, since
the charged gluons are presumably confined, and therefore only contribute
to the self-energy of each quark.  The string tension of the
$<(0|..|0)>$ term should therefore be zero in an abelian-projection
gauge; there should be no area suppression whatever, in any range
of quark separations.  This is an easy prediction to test.

   Let $U(C)$ represent a product of link variables along the path $C$.
In $SU(2)$ this can always be expressed as

\beq
       U(C) = a_0 {\bf 1} + i a_k \s^k
\eeq
where the $\s^k$ are the Pauli matrices.  Then, in the adjoint
representation

\bea
    U_{++}(C) \equiv   (1|U_A(C)|1) &=& a_0^2 - a_3^2 + 2i a_0 a_3
\non \\
    U_{--}(C) \equiv  (-1|U_A(C)|-1) &=& a_0^2 - a_3^2 - 2i a_0 a_3
\non \\
    U_{00}(C) \equiv   (0|U_A(C)|0) &=& 2(a_0^2 + a_3^2) - 1
\eea
Define the charged and neutral Wilson loops

\bea
        W_c(C) &=& <U_{++}(C)> = <U_{--}(C)>
\non \\
        W_0(C) &=& <U_{00}(C)>
\eea
together with the corresponding Creutz ratios $\chi_c[R,T]$ and
$\chi_0[R,T]$.  As discussed above, the abelian projection prediction
is that $\chi_0[R,T] = 0$ (or, at least, that \break
$\chi_0 << \chi_c$) in an abelian projection gauge.

   We have computed these Creutz ratios by lattice Monte Carlo,
in D=3 dimensions with a Wilson action at $\beta=5$, which is just
inside the D=3 scaling region.  The charged and neutral loops
were evaluated in maximal abelian gauge.  The results obtained
after 153000 update iterations, for $\chi_0[R,R]$, $\chi_c[R,R]$,
and for $\chi[R,R]$ (the Creutz ratio of gauge-invariant
adjoint $j=1$ loops), are shown in Fig. 6.
Creutz ratios for the fundamental ($j=\oh$) loop $\chi_F(R,R)$ are also
displayed in Fig. 6.  It can be seen that Creutz ratios in the
$j=\oh$ and $j=1$ representations are in the proportion predicted by
large-N, which is a ratio of Casimirs

\beq
          {\chi[R,R] \over \chi_F[R,R]} \approx {8 \over 3}
\eeq

  It can also be seen that $\chi_0$ is not zero, and shows no tendency
to go zero faster than $\chi_c$.  In fact, although $\chi_0$ is
smaller than $\chi_c$, the difference is only about 10\%; this difference
does not appear to grow with loop size.  Thus, while
there may perhaps be some effect attributable to abelian monopoles,
the effect seems rather small; certainly it is not sufficient to
explain confinement in this region.  This result seems entirely
consistent with the results in ref. \cite{Win}.

   We have also studied the correlation of Polyakov lines
in D=4 dimensions and maximal abelian gauge
\bea
       P_0(R) &=& <U_{00}[L_1] U_{00}[L_2]>
\non \\
       P_c(R) &=& <U_{++}[L_1] U_{--}[L_2]>
\eea
where $L_{1,2}$ are parallel Polyakov loops, 2 lattice spacings in length,
separated by R lattice spacings in the spatial hyperplane.  Because of
the small time extension, we must work at a strong coupling - in our
case $\beta=1.8$ - to avoid the deconfinement transition.  Although this
is at strong-coupling, it is at least close to the strong-to-weak
coupling transition point, and we can see if there is any tendency for
the "neutral" correllations $P_0(R)$ to behave differently from the
"charged" correlations $P_c(R)$.  The data is shown in Fig. 7.
Charge screening sets in at about 2 lattice spacings, and there
does not appear to be any difference in the behavior of $P_0$ and $P_c$.

For the numerical simulations, we used an heat-bath algorithm, running
on a $16^3~\times~2$ lattice at $\beta=1.80$. The results were obtained
averaging over 1000 configurations, after 20000 sweeps of thermalization.
The continous lines in Fig. 7 are the results of fit for the
thermodynamical mixing of two interaction channels as described in ref.
\cite{MM-Beirl}. For the first Boltzmann factor we used the lattice version
of a Coulomb plus linear potential plus self energy and via the second term
we took the screening of charges into account.

   The conclusion of this numerical work is that there seems to be
no appreciable difference in the forces between abelian charged and
abelian neutral adjoint quarks, at the couplings and separations
we have investigated.  This evidence does not favor the abelian
projection theory.

   It may be appropriate, at this point, to comment on certain other
Monte Carlo investigations of the abelian projection theory.
Numerical work on this problem was initiated by Kronfeld et. al
\cite{Kronfeld}, who found a drop in the monopole density at
the deconfinement phase transition, in "maximal abelian gauge"
defined as the gauge which maximizes the quantity

\beq
     Q = \sum_x \sum_{\m=1}^{4} Tr[\s_3 U_\m(x) \s_3 U^{\dagger}_{\m}(x)]
\eeq
  More recent
investigations by Del Debbio et. al. have shown, however, that
the definition of monopole density is plagued by lattice artifacts,
and is not at all a good order parameter for confinement \cite{Luigi}.
For example, it is found the monopole density neither shows correllation
with the string tension with cooling, nor a drop across the
deconfinement phase transition.  These problems may be alleviated by a
more appropriate definition of the monopole creation operator, but we
will not pursue that issue here.

   An alternate line of investigation is to see if the electric
flux inside the flux tube is dominated by the Cartan subalgebra; e.g.
in SU(2), with the choice of maximal abelian gauge above, one
checks to see if the field-strength is proportional to $\s_3$.
Let $U_{\m\n}$ be the plaquette variable.  One defines the field
strength on the lattice as $F^{a}_{\m\n} = Tr[U_{\m\n}-U^{\dg}_{\m\n}]
\s^a]/4i$, as well as the quantities

\beq
             J^a = {<D^a TrW> \over <TrW>} - <D^a>
\label{Ja}
\eeq
where  $W$ is a Wilson loop in the $\m \n$ plane; and
$D^a = \sum (F^a_{\m \n}(x))^2/n_P$, where the sum is over
central plaquettes in the minimal area bounded by the loop,
furthest from the boundary.  $n_P$ is the number of central plaquettes.
Let $J=J^1+J^2+J^3$, and define the ratio

\beq
         \rho = {J^3 \over J}
\eeq
If $\rho \approx 1$ this would tend to confirm the abelian projection
theory, while $\rho \approx 1/3$ would tend to refute it.  The result
found in \cite{Win,Olejnik} was $\rho \approx 1/3$.

   On the other hand, Hioki et. al. \cite{Hioki} have argued that
this ratio should be
defined using abelian loops in the numerator.  An abelian link
is defined as the diagonal part of the link variable, rescaled to
restore unitarity.  An abelian loop is a loop constructed from
abelian links. Let $u_{\m\n}$ be a $1 \times 1$ abelian loop, and
define

\bea
        \rho_A &=& {J_A \over J}
\nonumber \\
        J_A &=& {<{1 \over n_P}\sum f_{\m\n}^2 TrW> \over <TrW>} -
<f_{\m\n}^2>
\non \\
        f_{\m \n} &=& (u_{\m\n}-u_{\m\n}^*)/2i
\label{Jabel}
\eea
With these definitions, Hioki et. al. find $\rho_A$ actually greater
than 1 (up to $\rho_A \approx 1.6$, in fact).

   We regard conclusions based on the enhancement of abelian
loops, both in ref. \cite{Hioki} and also in \cite{Yots} as very
misleading for the following reason:  Maximal abelian gauge
simply makes links as diagonal as possible; the enhancement of
abelian loops (termed "abelian dominance") which was found in
\cite{Yots} is a simple consequence of this
fact. It is therefore crucial to choose observables whose behavior
will really test the abelian projection theory, rather than just
display this particular aspect of the gauge condition.  The behavior
of abelian Wilson loops,
and the $\rho_A$ quantity defined above, do not meet that criterion.

   The following calculation will illustrate the point.
Instead of gauge-fixing to maximal abelian gauge, let us fix to
another, "x-y" maximal abelian gauge, introduced in ref. \cite{Win},
which is defined to maximize the quantity

\beq
     Q = \sum_x \sum_{\m=1}^{2} Tr[\s_3 U_\m(x) \s_3 U^{\dagger}_{\m}(x)]
\eeq

\noindent This gauge forces links in the x and y directions only
to be as diagonal as possible.
Since there is no requirement, in the abelian projection theory,
that the gauge-fixing condition must
be spherically symmetric, this gauge should be as good as the usual
maximal abelian gauge.
We now compute $\rho_A$ by Monte Carlo separately for loops
oriented in the x-y, and z-t planes.  The results have only been computed
for rather small loops, but we feel they already show how things go.
The loops are from R=1 to R=4 lattice spacings wide, and T=4 lattice
spacings high.  Half of the plaquettes in the minimal area of
the loop were used for calculating $J_A$; these are the $2 \times R$
plaquettes which are one lattice spacing away from loop boundaries
in the $T$ direction.

    The results of this calculation, performed in D=4 dimensions at
$\beta=2.4$, are shown in Fig. 8. If one accepts the proposition that
$\rho_A \approx 1$ is evidence for the abelian projection theory, then
it would seem from this data that monopoles
are responsible for confinement in the x-y plane, but not in the z-t plane.
That conclusion, we feel, is nonsense.  A much more reasonable explanation
is that, because links are nearly diagonal in the x-y plane, loops
which are built from the diagonal part of the links pretty well
approximate the full Wilson loop.  Conversely, in the z-t plane, the
results look much as if there were no gauge-fixing at all.

   As an additional check, we have also computed $\rho_A$ for a
small $R=2$, $T=4$ loop at $\beta=2.8$, in ordinary maximal
abelian gauge.  At this value
of $\beta$, for such a small loop, there should be no flux tube formed;
yet we find $\rho_A = 0.88$.  This is another indication that the large
value of $\rho_A$ is a gauge effect, which has nothing to do with
flux tube formation.

   This calculation illustrates the fact that, in maximal abelian
gauge, one must ensure that the observables chosen are relatively
insensitive to the diagonality of links which is enforced by the gauge.
This is the case for the $J^a$ observables of eq. \rf{Ja}, whose sum is
directly related to the energy density in the flux tube; it is not
the case for the $J_A$ observable of eq. \rf{Jabel}, which is not related
directly to the energy density.

\section{Conclusions}

   A prediction of the abelian projection theory of confinement is
that, in an abelian projection gauge, adjoint quark colors which are
neutral with respect to the remnant $U(1)^{N-1}$ symmetry are
oblivious to the confining force.  This prediction turns out to conflict
with large-N arguments, and can be tested by looking for the
{\it absence} of an area-law term, over any length scale, in Wilson loops
and Polyakov lines with appropriate insertions of abelian-neutral
projection operators.  We have found, instead, that insertion of
abelian neutral (and abelian charged) projection operators has
very little effect on the value of the string tension extracted
from the loop.  Assuming that the QCD flux tube in abelian projection
gauge is abelian neutral (since it is supposed to be formed by
the "photon" fields), this means that there is no significant
difference in the forces between abelian neutral adjoint quarks,
and between abelian charged adjoint quarks.  Thus our data is consistent
with large-N expectations, and in conflict with the abelian-projection
theory, which holds that the confining force is sensitive mainly to
the $U(1)^{N-1}$ charge.

    We have also examined the claim of "abelian dominance" in
maximal abelian gauge, found in \cite{Hioki,Yots}.  We believe
that our data in an "x-y maximal abelian gauge" (together with
previous work along these lines in ref. \cite{Win}), demonstrates that
this enhancement of abelian loops is purely a gauge effect, with
no relevance at all to the physics of confinement.

    Finally, we have calculated the force between QCD flux tubes
in the lattice strong-coupling expansion, and find that this
force is attractive.  If the strong-coupling result survives
in the continuum theory (and there is reason, based on the general
notion of flux coalescence, to think that it might), then QCD
flux tubes are {\it not} analogous to Abrikosov vortices in
type II superconductors, which tend to repel one another.

    In general, theories of confinement which rely on analogies
to abelian gauge theories will tend to identify a small subset
of the degrees of freedom, e.g. associated with
the $Z_N$ center or the Cartan
subgroup of the full gauge group, as being especially important
for quark confinement.  Such a subset becomes negligibly small compared to
the total number of degrees of freedom in the $N \ra \infty$ limit,
and it is therefore not surprising that such theories will somewhere
contradict results based on large-N counting arguments.  One theory
of confinement which is based rather explicitly on the large-N
picture is the "gluon-chain" model of flux-tube formation, advocated by
one of us in ref. \cite{gchain1}.  This model is consistent with
all large-N predictions, and also has some numerical support
\cite{gchain2}.  It is relevant here as an example of a confinement
mechanism in non-abelian gauge theories which has no abelian
counterpart; we believe that this must be true of any mechanism
which is consistent with large-N results.

\vspace{20pt}

\noindent {\Large \bf Acknowlegments}{\vspace{11pt}

   We thank the organizers of the Hadron Structure 91 meeting held in
Stara Lesna, Czechoslovakia, where this work was begun.  We also thank
J. Ho\v{s}ek for drawing our attention to the relevance of vortex repulsion
to dual-superconductor models.  M.F. is grateful to R. Dirl
and P. Kasperkovitz for valuable suggestions concerning group-theoretical
aspects of this investigation; J.G. acknowledges the hospitality of the
Niels Bohr Institute and the Lawrence Berkeley Laboratory.  The computer
calculations were carried
out in Tallahassee, Florida and Pisa, Italy; the Florida portion was
supported by the Florida State University Supercomputer Computations
Research Institute, which is partially funded by the U.S. Department
of Energy through Contract No. DE-FC05-85ER250000.

\newpage

\noindent {\Large \bf Figure Captions}
\bigskip
\bigskip

\begin{description}
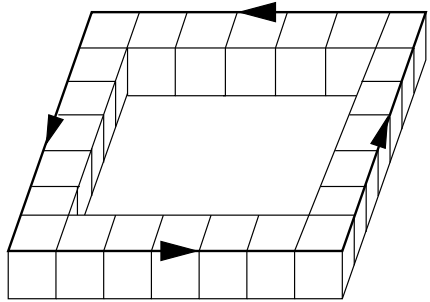
\item[Fig. 1] Leading strong-coupling contributions to the adoint
Wilson loop. (a) the $O(N^2)$ area contribution; (b) the $O(1)$
perimeter contribution.
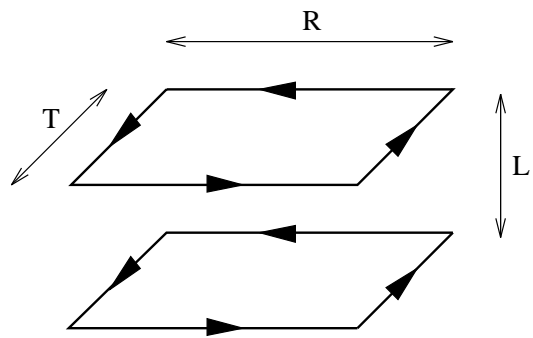
\item[Fig. 2] Two parallel quark-antiquark flux tubes realized
by two parallel Wilson loops of the same orientation.
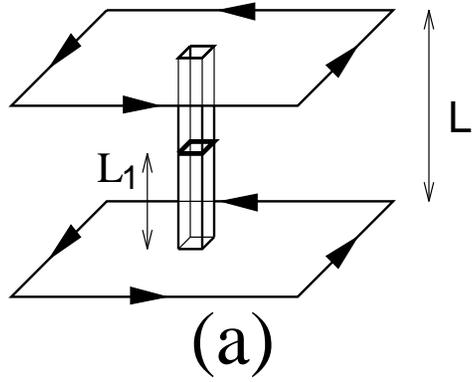
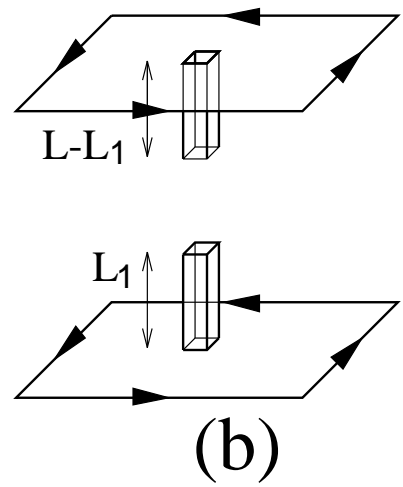
\item[Fig.3] Diagrams used in computing flux-tube interactions (eq.
\rf{A}): (a) term $A_a$, (a plaquette in the middle of a tube);
(b) term $A_b$.  It is necessary to sum over the position of the tube
in the sheet, and over the position of the "middle" plaquette along the
tube.
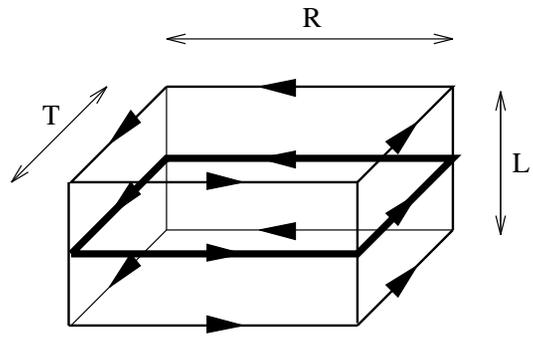
\item[Fig. 4] Leading contribution for parallel Wilson loops at
$L<<R$.  The sheet in the middle (bounded by the heavy solid line)
is in the fundamental representation of N-ality $N-2$.
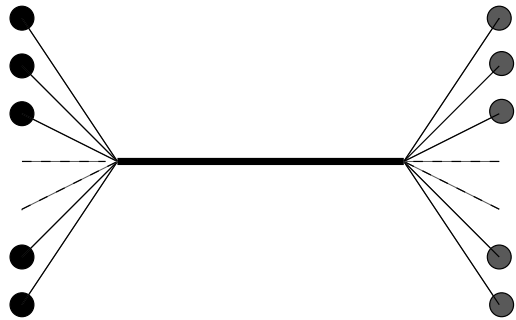
\item[Fig. 5] Coalescence of $n$ quark-antiquark flux tubes into a single
flux tube of N-ality $(N-n)\mbox{mod}N$.
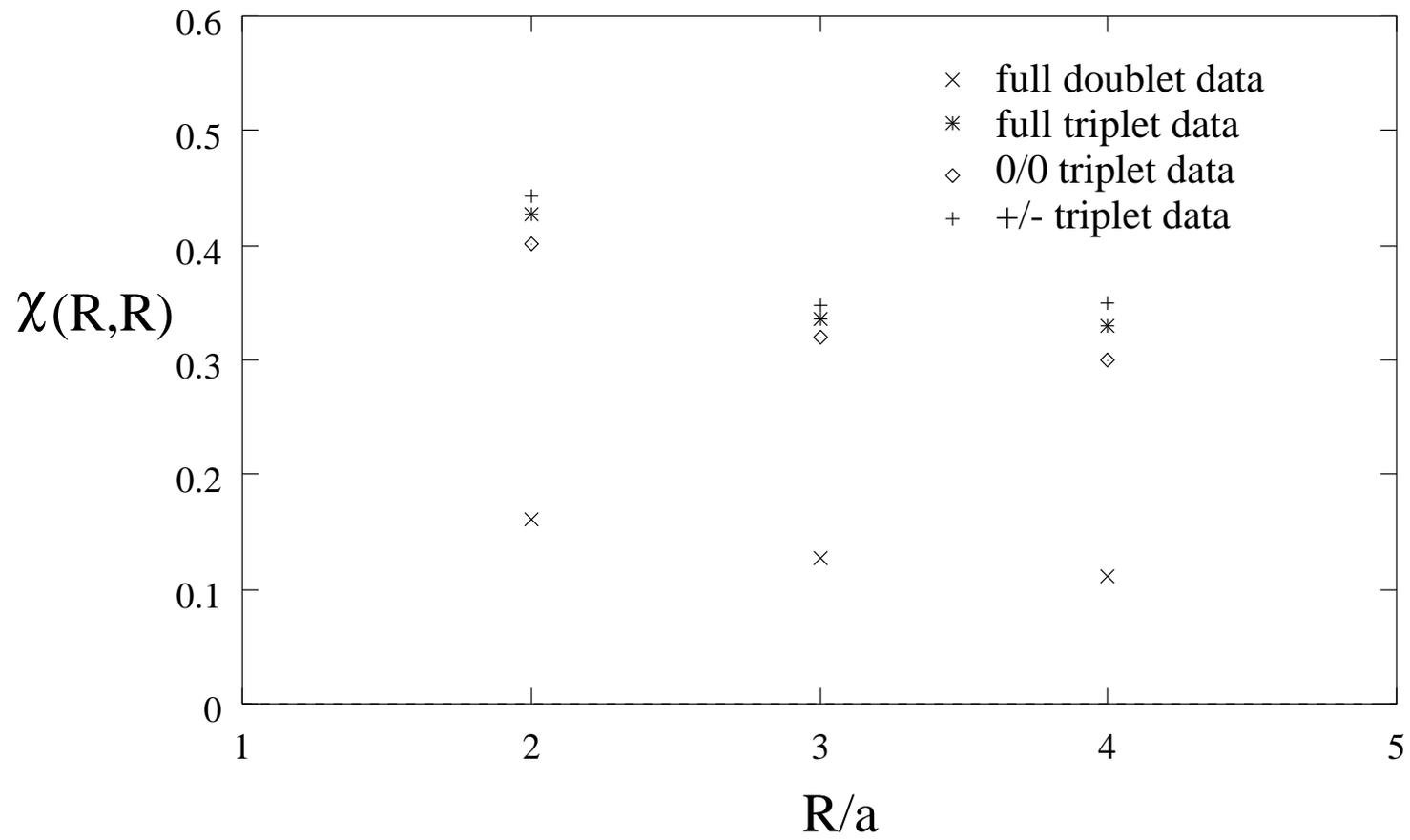
\item[Fig. 6] Creutz ratios of adjoint ($j=1$)Wilson loops. Stars are the
usual $\chi$ ratios, diamonds are ratios $\chi_0$ of loops
with abelian-neutral ($m=0$) projection operators; vertical crosses are
ratios
$\chi_c$ of loops with abelian-charged ($m=\pm 1$) projection
operators, evaluated in maximal abelian gauge.  The errors on loops at
$R=1,2,3a$ are negligible, errors at $R=4a$ are about $\pm 10 \%$.
Also shown (by diagonal crosses) is $\chi_F$, the usual Creutz
ratio in the fundamental ($j=\oh$) representation.  Simulation is in
D=3 dimensions at $\beta=5$.
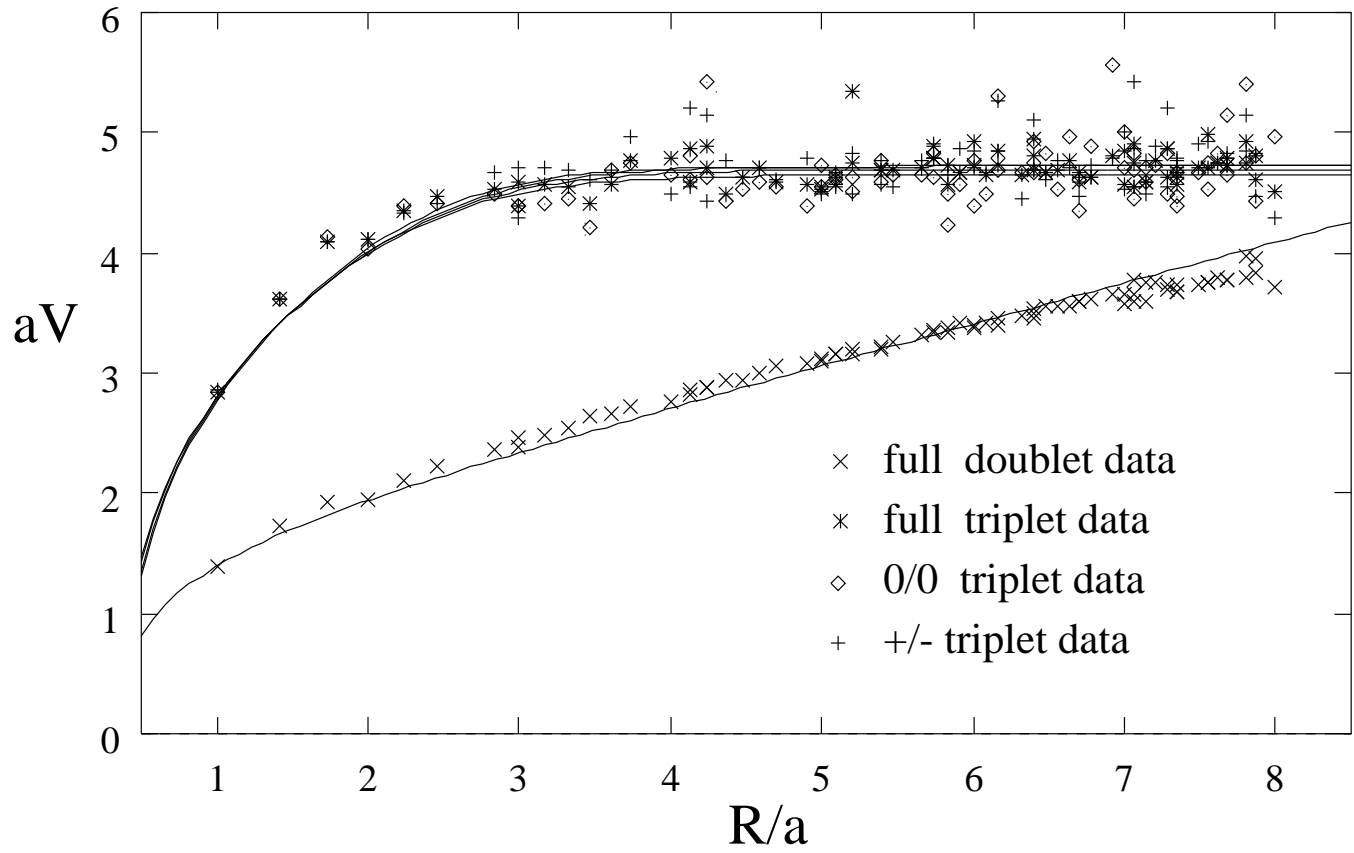
\item[Fig. 7] The potential $V(R)$ extracted from Polyakov lines
of extension $N_t=2$ lattice spacings. The potentials for "neutral"
(diamonds) and "charged" (vertical crosses) sources in maximal abelian
gauge of SU(2) are compared with the ungauged potentials (stars) in the
adjoint representation. For comparison we show also the potential between
doublet sources. The statistical errors start to get larger than the
symbols in the screening region and can be estimated from the fluctuations
with the distance R. The full lines are the results of fits with a sum of
two Boltzmann factors.
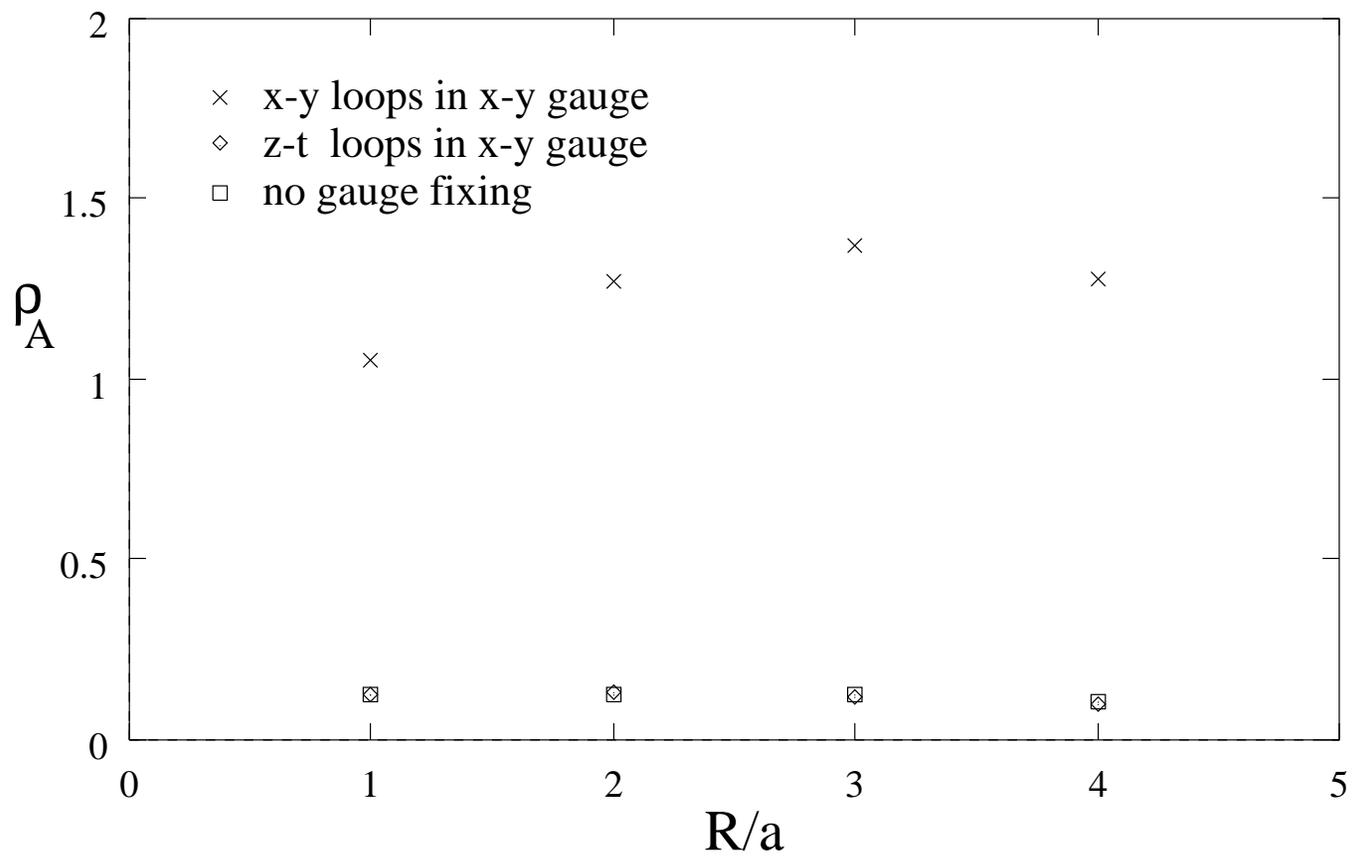
\item[Fig. 8] $\rho_A$ in the "x-y maximal abelian" gauge.  Crosses
represent $\rho_A$ extracted from loops in the x-y plane; diamonds
represent $\rho_A$ extracted from loops in the z-t plane.  Squares
are values of $\rho_A$ taken with no gauge fixing.

\end{description}


\vspace{33pt}

\begin{thebibliography}{xx}
\bibitem{fluxon} G. 't Hooft, Nucl. Phys. B138 (1978) 1;
G. Mack, in "Recent Developments in Gauge Theories," Proceedings
of the 1979 Cargese Summer Institute, ed. G 't Hooft et. al.
(Plenum, New York, 1980).
\bibitem{GH} J. Greensite and M. Halpern, Phys. Rev. D27 (1983) 2545.
\bibitem{MC} M. Faber, W. Kleinert, M. M\"uller, and S. Olejnik, in
"Hadron Structure 91", Proceedings of the HS91 Conference in Stara
Lesna, Czechoslovakia, (Slovak Acad. of Sci., Bratislava,1991);
H. Faber and H. Markum, Nucl. Phys. B (Proc. Suppl.) 4 (1988), 204;
C. Michael, Nucl. Phys. B259 (1985) 58;
J. Ambjorn, P. Olesen, and C. Peterson, Nucl. Phys. B240 [FS12]
(1984) 533.
\bibitem{'t Hooft} G. 't Hooft, Nucl. Phys. B190 [FS3] (1981) 455.
\bibitem{Kronfeld} A. Kronfeld, M. Laursen, G. Schierholz, and
U.-J. Wiese, Nucl. Phys. B293 (1987) 461;
\bibitem{Rosenzweig} C. Rosenzweig, Phys. Rev. D38 (1988) 1934.
\bibitem{Luigi} L. Del Debbio, A. Di Giacomo, M. Maggiore, and
S. Olejnik, Phys. Lett. B267 (1991) 254.
\bibitem{Win} J. Greensite and J. Iwasaki, Phys. Lett. B255 (1991) 415;
J. Greensite and J. Winchester, Phys. Rev. D40 (1989) 4167.
\bibitem{Olejnik} A. Di Giacomo, M. Maggiore, and S. Olejnik, Nucl.
Phys. B347 (1990) 441.
\bibitem{Hioki} S. Hioki, S. Kitahara, S. Kiura, Y. Matsubara,
O. Miyamura, S. Ohno and T. Suzuki, Phys. Lett. B272 (1991) 326.
\bibitem{Yots} T. Suzuki and I. Yotsuyanagi, Phys. Rev.
D42 (1990) 4257.
\bibitem{Smit} J. Smit and A. van der Sijs, Nucl. Phys. B355 (1991) 603.
\bibitem{MM-Beirl} M. M\"uller, W. Beirl, M. Faber, H. Markum, Nucl. Phys.
B (Proc.Suppl.) 26 (1992) 423.
\bibitem{gchain1} J. Greensite, Nucl. Phys. B249 (1985) 263.
\bibitem{gchain2} J. Greensite, Nucl. Phys. B315 (1989) 663.
\end{thebibliography}
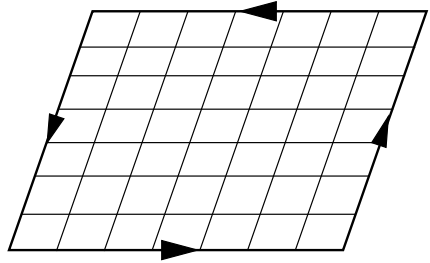
\end{document}